\documentclass[review]{elsarticle}

\usepackage{mathtools}  

\usepackage{hyperref}
\usepackage{cleveref}
\hypersetup{pdfborder=0 0 0}
\usepackage[hyphens]{xurl}

\usepackage{graphicx}
\usepackage{caption}
\usepackage{subcaption}
\usepackage{textcomp}
\usepackage{xcolor}
\usepackage{colortbl}
\usepackage{booktabs}
\usepackage{tabularx}
\usepackage{adjustbox}
\usepackage{soul}
\usepackage{color}
\usepackage{tcolorbox}

\usepackage{enumitem,amssymb}
\newlist{todolist}{itemize}{2}
\setlist[todolist]{label=$\square$}
\newlist{avoidlist}{itemize}{2}
\setlist[avoidlist]{label=$\bigtimes$}

\definecolor{archived-blue}{rgb}{0,0,1}
\definecolor{os-green}{rgb}{0.416, 0.66, 0.31}
\definecolor{reachable-yellow}{rgb}{0.945,0.76,0.196}
\definecolor{uponrequest-brown}{rgb}{0.75,0.565,0}
\definecolor{broken-orange}{rgb}{0.9, 0.57, 0.22}
\definecolor{unavailable-red}{rgb}{0.6,0,0}
\definecolor{private-grey}{rgb}{0.72,0.72,0.72}
\definecolor{proprietary-grey}{rgb}{0.26,0.26,0.26}

\newenvironment{highlightbox}[1]{
    \begin{tcolorbox}[title={#1}]
    }{
    \end{tcolorbox}
}

\definecolor{hlrevmeta}{RGB}{163, 255, 247}
\DeclareRobustCommand{\revmeta}[1]{#1}

\journal{Journal of Systems and Software}

\begin{document}

\newcommand\nauthors{40} 
\newcommand\nauthorsnew{5} 
\newcommand\requestcompleted{21} 
\newcommand\requestcompletedperc{52.5} 
\newcommand\requestanswered{30} 
\newcommand\requestansweredperc{75.0} 
\newcommand\requestunanswered{8} 
\newcommand\requestunansweredperc{20.0} 
\newcommand\requestundeliverable{2} 
\newcommand\requestundeliverableperc{5.0} 
\newcommand\daysresponse{23.8} 
\newcommand\dayscomplete{29.25} 
\newcommand\datasetsimproved{10} 
\newcommand\datasetsimprovedperc{17.5} 
\newcommand\datasetsclarified{30} 
\newcommand\datasetsclarifiedperc{52.6} 
\newcommand\datasetsaddressed{40} 
\newcommand\datasetsaddressedperc{70.1} 
\newcommand\datasetsarchived{8} 
\newcommand\datasetsarchivedperc{14.0} 
\newcommand\datasetsavailable{15} 
\newcommand\datasetsavailableperc{26.3} 
\newcommand\datasetslost{19} 
\newcommand\datasetssensitive{11} 
\newcommand\implementationsimproved{7} 
\newcommand\implementationsimprovedperc{19.4} 
\newcommand\implementationsclarified{10} 
\newcommand\implementationsclarifiedperc{27.7} 
\newcommand\implementationsaddressed{17} 
\newcommand\implementationsaddressedperc{47.2} 
\newcommand\implementationsarchived{6} 
\newcommand\implementationsarchivedperc{16.6} 
\newcommand\implementationsavailable{11} 
\newcommand\implementationsavailableperc{30.5} 
\newcommand\implementationslost{6} 
\newcommand\implementationsproprietary{4} 

\begin{frontmatter}

\title{Requirements Quality Research Artifacts: Recovery, Analysis, and Management Guideline}

\author[bth]{Julian Frattini}
\cortext[mycorrespondingauthor]{Corresponding author}
\ead{julian.frattini@bth.se}

\author[uh]{Lloyd Montgomery}
\ead{lloyd.montgomery@uni-hamburg.de}
\ead[URL]{https://lloydm.io/}

\author[bth]{Davide Fucci}
\ead{davide.fucci@bth.se}
\ead[URL]{https://dfucci.github.io/}

\author[bth]{Michael Unterkalmsteiner}
\ead{michael.unterkalmsteiner@bth.se}
\ead[URL]{https://www.lmsteiner.com/}

\author[bth,fortiss]{Daniel Mendez}
\ead{daniel.mendez@bth.se}
\ead[URL]{https://www.mendezfe.org/}

\author[fortiss,netlight]{Jannik Fischbach}
\ead{jannik.fischbach@netlight.com}

\address[bth]{Blekinge Institute of Technology, Valhallavägen 1, 371 41 Karlskrona, Sweden}
\address[uh]{University of Hamburg, 20146 Hamburg, Germany}
\address[netlight]{Netlight Consulting GmbH, Sternstraße 5, 80538 Munich, Germany}
\address[fortiss]{fortiss GmbH, Guerickestraße 25, 80805 Munich, Germany}

\begin{abstract}
    Requirements quality research, which is dedicated to assessing and improving the quality of requirements specifications, is dependent on research artifacts like data sets (containing information about quality defects) and implementations (automatically detecting and removing these defects). 
    However, recent research exposed that the majority of these research artifacts have become unavailable or have never been disclosed, which inhibits progress in the research domain. 
    In this work, we aim to improve the availability of research artifacts in requirements quality research. 
    To this end, we (1) extend an artifact recovery initiative, (2) empirically evaluate the reasons for artifact unavailability using Bayesian data analysis, and (3) compile a concise guideline for open science artifact disclosure.
    Our results include \datasetsimproved\ recovered data sets and \implementationsimproved\ recovered implementations, empirical support for artifact availability improving over time and the positive effect of public hosting services, and a pragmatic artifact management guideline open for community comments. 
    With this work, we hope to encourage and support adherence to open science principles and improve the availability of research artifacts for the requirements research quality community.
\end{abstract}


\end{frontmatter}


\section{Introduction}
\label{sec:intro}

Requirements quality research is a sub-domain of requirements engineering research specifically focused on assessing and improving the quality of requirements specifications~\cite{montgomery2022empirical}. Requirements quality research depends on research artifacts: annotated \textit{data sets} are used as the ground truth about quality violations, and \textit{implementations} are deliverable artifacts (e.g., tools) that are often intended to be transferred into industry in order to support the management of requirements quality.

However, recent research exposed that the majority of research artifacts are not available anymore or have never been~\cite{montgomery2022empirical,frattini2022live,zhao2021natural}. This inhibits the progress of the requirements quality research domain, as new contributions cannot reuse existing data sets for benchmarking their approach or evolve existing implementations.
Instead, they have to resort to recreating already proposed solutions.

In this work, we aim to address the problem both retrospectively (i.e., recovering unavailable artifacts of prior publications) and prospectively (i.e., offering guidance for future publications). To this end, we make the following contributions:

\begin{enumerate}
    \item Artifact recovery (C1): a two-phase recovery initiative of previously unavailable research artifacts, resulting in \datasetsimproved\ recovered data sets and \implementationsimproved\ recovered implementations. This improves the availability of research artifacts in the requirements quality research domain, recovering lost opportunities for the reproduction of empirical results or reuse of developed tools.
    \item Evaluation (C2): an empirical evaluation of the reasons for artifact unavailability using Bayesian data analysis. The results help to better understand the potential barriers in disclosing artifacts and steer future open science initiatives within the requirements quality research community.
    \item Open Science Artifact Management Guideline (C3): a concise and pragmatic guideline summarizing recommendations on how to collect, document, license, archive, and share research artifacts. The guideline supports authors of scientific work in disclosing research artifacts to maintain their availability and increase their impact.
\end{enumerate}

An initial version of the first contribution---the first of the two phases of the artifact recovery initiative---has been published in the \textit{Natural Language Processing for Requirements Engineering} (NLP4RE) workshop\footnote{\url{https://nlp4re.github.io/2023/}}~\cite{frattini2023let}. In this paper, we extend the artifact recovery initiative by a second phase (C1) and add two more contributions (C2 and C3) to deepen the insights from the gathered data and provide actionable guidance for future publications. We reused minor parts of the original manuscript in a verbatim manner, such as the presentation of related work or terminological definitions.

The rest of this manuscript is structured as follows: \Cref{sec:background} introduces the background on both the requirements quality research domain and open science principles. \Cref{sec:recovery} describes the two-phase initiative to recover unavailable research artifacts, and \Cref{sec:eval} empirically evaluates the gathered data to infer reasons for the unavailability of these artifacts. Finally, \Cref{sec:guide} presents a concise guideline on artifact management, offering support in preserving artifact availability, before concluding the paper in \Cref{sec:conclusion}. 

\vspace{-0.3cm}

\subsection*{Data Availability Statement}

All study material, including our data, code, and evaluation reports, are accessible in our replication package, archived at \url{https://doi.org/10.5281/zenodo.7708570}. The guideline presented in \Cref{sec:guide} is archived at \url{https://doi.org/10.5281/zenodo.8134402} and a collaborative version is accessible at \url{https://bit.ly/OSAMG}. 

\section{Background}
\label{sec:background}

\subsection{Open Science in Software Engineering}
\label{sec:background:os}

A vital property of scientific work is reproducibility~\cite{mendez2020open,feitelson2015repeatability}, i.e., the ability to ``duplicate the results of a prior study using the same material as were used by the original investigator''~\cite{cacioppo2015social}, as it strengthens the robustness of scientific findings contributed to a field of research~\cite{anda2008variability,mcnutt2014reproducibility,cacioppo2015social}. One necessary precondition of reproducibility is the availability of study materials, including study protocols to reproduce an investigation, data to re-evaluate statistical claims, or source code to recreate tools. To emphasize this dependency, Minocher et al. introduced a four-stage model of reproducibility~\cite{minocher2020reproducibility} consisting of \textit{data recoveribility}, \textit{data usability}, \textit{analytical clarity}, and \textit{agreement of results}. The model is sequential, i.e., \textit{analytical clarity} of data is meaningless if the data is not \textit{recoverable}. Hence, all reproducibility hinges on the availability or recoverability of research artifacts.
Furthermore, the model applies to all types of research.
While the \textit{agreement of results} is more straightforward to obtain for studies involving quantitative data, studies dealing with qualitative data should at least allow an external reviewer to review the analysis process and understand how the authors arrived at their conclusions.

\textit{Open science} is an initiative dedicated to ensuring public availability of research artifacts~\cite{mendez2020open}. Within open science, the facets of \textit{open access} for publications, \textit{open data} for data sets, and \textit{open source} for source code are most relevant to software engineering~\cite{tennant2019foundations}, where each facet of open science entails different techniques and best practices to disclose its respective type of research artifacts. Several governmental research funding agencies, including ones of the European Union, made open access to scientific results (including data, tools, etc.) mandatory.\footnote{\url{https://research-and-innovation.ec.europa.eu/strategy/strategy-2020-2024/our-digital-future/open-science/open-access_en}}


Open science entails its own set of challenges.
Most notably, adherence to open science principles requires additional effort in light of documenting and disseminating research artifacts~\cite{winter2022retrospective} and is sometimes even impossible, as sensitive data may not be shareable~\cite{LTG20}.
However, because of the importance of replication and reproduction for a scientific field~\cite{shepperd2018role} and the contribution of open science toward these properties~\cite{mendez2020open}, the software engineering research community has come to the agreement that open science is not only worth but rather imperative to pursue~\cite{krishnamurthi2015real}.

Recent endeavors to incentivize scholars to follow open science principles include open science badges~\cite{kidwell2016badges,krishnamurthi2015real,hermann2020community} and registered reports~\cite{nosek2018preregistration}. 
The introduction of artifact evaluation tracks at premiere SE conferences like ESEC/FSE\footnote{\url{https://conf.researchr.org/series/fse}} greatly benefited the availability of research artifacts in SE research~\cite{krishnamurthi2015real}.
However, several sub-domains of the SE research community are still in the process of adapting and implementing open science principles~\cite{mendez2020open}, and the unavailability of artifacts remains common in areas like requirements quality research~\cite{montgomery2022empirical,zhao2021natural}.
Prominent reasons for the unavailability of artifacts include the sensitivity of data~\cite{krishnamurthi2015real}, corresponding authors changing their affiliation and consequently losing access to their artifacts~\cite{gabelica2022many,winter2022retrospective}, or authors not seeing any benefit in sharing their artifacts~\cite{wacharamanotham2020transparency}.
While some reasons for the unavailability of artifacts (e.g., the sensitivity of company-owned data) may well require significant effort to cope with or are unavoidable, other reasons (e.g., loss of artifact, lack of diligence) can be circumvented easily by following guidelines~\cite{mendez2020open} and making use of modern tools, e.g., Zenodo, for artifact sharing~\cite{sicilia2017community}.

\subsection{Requirements Quality Literature}
\label{sec:background:rq}

Artifacts produced in requirements engineering (RE)---e.g., systematic requirements specifications, use cases, or user stories---have a significant impact on downstream software development activities~\cite{wagner2019status}, potentially causing project delay or even failure~\cite{fernandez2013naming}. Consequently, requirements artifacts merit quality assurance~\cite{montgomery2022empirical}. The requirements quality literature is dedicated to providing an understanding as well as the support for measuring and improving the quality of requirements~\cite{montgomery2022empirical}. One popular approach to this is the proposal of \textit{quality factors}~\cite{frattini2022live}. Requirements quality publications often propose one or more quality factors---e.g., the use of \textit{coordination ambiguity} leading to divergent interpretations~\cite{ezzini2021using} or the use of \textit{passive voice} causing the omission of information~\cite{femmer2014impact}---, provide a \textit{data set} where instances of violations against that quality factor are annotated, and finally present an \textit{implementation} (i.e., an algorithm or full-fledged tool) to detect or remove these instances automatically.

These artifacts---both data sets and implementations---represent essential contributions facilitating empirical research and technology transfer. While the (annotated) data sets are the main drivers for developing new and improving existing implementations for quality factor detection, implementations are the research deliverables to be deployed in industry for actual integration and improvement of the software engineering process~\cite{krishnamurthi2015real}. 

However, the degree to which artifacts are disclosed in the requirements quality literature varies~\cite{montgomery2022empirical}. The majority of research artifacts are simply \textit{unavailable}, i.e., authors present them in a publication but provide no access to them~\cite{zhao2021natural}. \textit{Proprietary} artifacts, i.e., those turned into a commercial product, or \textit{private} artifacts, i.e., those not disclosed due to sensitivity, constitute an excused exception to this group. Among those artifacts that are actually disclosed, i.e., hosted in an accessible way and referred to from the manuscript, several are \textit{broken} since the URL pointing toward the artifact does not resolve anymore. Other artifacts are only available \textit{upon request}, and their access depends on the author's correspondence and upholding the promise made in the article. Among those artifacts that are actually \textit{reachable} (i.e., referred to with a non-broken source link), implementations where the source code is hosted openly and contains an open source license (i.e., \textit{open source}) are most common. For smaller data sets, it is also common that they are \textit{available in the paper} themselves, e.g., in the form of a table in the manuscript. In the best case, authors have properly \textit{archived} their artifact, which implies (1) an immutable URL, (2) permanent hosting, and (3) unrestricted accessibility. Only very few services (e.g., Zenodo~\cite{sicilia2017community}) offer this level of hosting.

Recent systematic studies revealed that a significant amount of the artifacts presented in the requirements quality literature are not available\footnote{Where available means a status of \textit{Upon request} (see \Cref{tab:statuscodes}) or better.} anymore or have never been~\cite{zhao2021natural,montgomery2022empirical,frattini2022live}. \Cref{tab:statuscodes} reports the availability status of 57 data sets (D) and 36 implementations (I) extracted from the 57 primary studies of a previously-published literature review on requirements quality factors~\cite{frattini2022live}. 
Most notably, the table visualizes that a large portion of data sets (87.8\%) and implementations (80.6\%) are not available anymore or never were (i.e., status \textit{broken} or worse).
A significant portion of data sets (42.1\%) are excused due to containing private or sensitive data, but for the majority of implementations (77.8\%), no valid excuse for their unavailability is provided.

\begin{table}
    \centering
    \footnotesize
    \caption{Availability status of requirements quality artifacts~\cite{frattini2022live}}
    \label{tab:statuscodes}
    \begin{tabularx}{\textwidth}{p{2cm}|X|rr|rr} \toprule
        \textbf{Status} & \textbf{Explanation} & \multicolumn{2}{c|}{\textbf{Datasets}} & \multicolumn{2}{c}{\textbf{Implementations}} \\ \hline
        \cellcolor{archived-blue} {\color{yellow} Archived} & The artifact is hosted in a service that satisfies the following criteria: (1) immutable URL (cannot be altered by the author or someone else), (2) permanent (the hosting organization has a mission to maintain artifacts for the foreseeable future), and (3) accessible (there is a DOI pointing to the real data source URL) & 1 & 1.7\% & 0 & {0\%} \\
        \cellcolor{os-green} Open Source & [only for implementations] The source code is disclosed and contains an open-source license (the artifact has a license which grants access and re-use) & - & & 5 & {13.9\%}  \\
        \cellcolor{os-green} Available in paper & [only for data sets] The data set is contained in the manuscript itself & 5 & {8.8\%} & - & \\
        \cellcolor{reachable-yellow} Reachable & The artifact is reachable now but missing some of the \textit{archived} or \textit{open source} aspects & 1 & {1.7\%} & 1 & {2.9\%} \\
        \cellcolor{uponrequest-brown} Upon request & Authors claim the artifact is available upon request & 0 & {0\%} & 1 & {2.9\%} \\ \hline
        \cellcolor{broken-orange} Broken & A link to the artifact is contained in the paper, but it does not resolve & 10 & {17.5\%} & 1 & {2.9\%} \\
        \cellcolor{unavailable-red} {\color{white} Unavailable} & An artifact is presented, but no indication on how to access it is provided & 15 & {26.3\%} & 27 & {77.8\%} \\
        \cellcolor{private-grey} Private & The authors state that an artifact exists but is private for some reasons (such as industry collaboration with private data, etc.) & 24 & {42.1\%} & 0 & {0\%} \\
        \cellcolor{proprietary-grey} {\color{white} Proprietary} & The artifact is proprietary, and access is granted upon payment & 1 & {1.7\%} & 1 & {2.9\%} \\ \hline
        Total & & 57 & & 36 \\
        \bottomrule
    \end{tabularx}
\end{table}

The visualization of the poor artifact availability~\cite{montgomery2022empirical,frattini2022live} shown in \Cref{tab:statuscodes} revealed that the requirements quality research domain suffers from a similar reproducibility crisis as other fields of research had~\cite{krishnamurthi2015real}.
While SE research, in general, has advanced over the last 10 years~\cite{winter2022retrospective,hermann2020community}, the requirements quality research domain suffers from a limited adoption of open science principles~\cite{mendez2020open}.
The consequently limited reproducibility of research in the field~\cite{minocher2020reproducibility} undermines the robustness of scientific findings~\cite{anda2008variability,mcnutt2014reproducibility,cacioppo2015social} and the technology transfer of scientific results into practice~\cite{gorschek2006model}.
There is a clear gap regarding the state of open science in the requirements quality research domain.

\section{Artifact Recovery}
\label{sec:recovery}

We aim to improve the \textit{availability} of research artifacts by (a) requesting authors of publications containing unavailable artifacts to recover their artifacts and (b) requesting authors of publications containing available but not archived artifacts to improve their artifacts' availability. 
To this end, we ask the following research question: to what degree can research artifacts from the requirements quality research domain be recovered by approaching their owners (RQ1)?
In \Cref{sec:recovery:design}, we document the design of the artifact recovery initiative. 
In \Cref{sec:recovery:result}, we present the results of this initiative.
\revmeta{In \mbox{\Cref{sec:recovery:threats}}, we discuss the threats to validity of these results.}

\subsection{Recovery Process}
\label{sec:recovery:design}

\begin{figure}[ht]
    \centering
    \includegraphics[width=\textwidth]{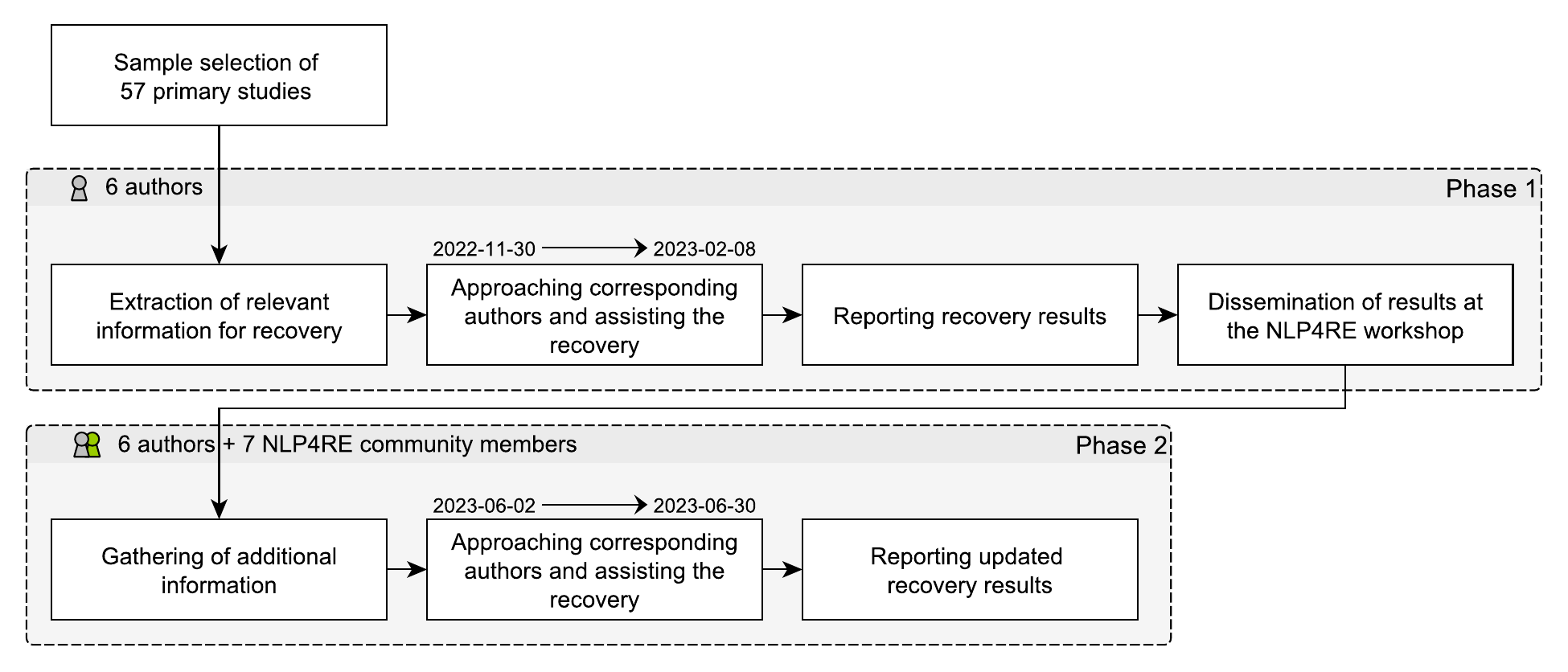}
    \caption{Overview of the two-phase recovery process}
    \label{fig:process}
\end{figure}

\Cref{sec:recovery:design:sample} describes the selection of primary studies from which we recover the artifacts. We detail our approach of contacting corresponding authors in \Cref{sec:recovery:design:approaching} and maintaining contact in \Cref{sec:recovery:design:correspondence}. \Cref{sec:recovery:design:evaluation} explains the evaluation of the collected data and \Cref{sec:recovery:design:crowdsourcing} documents the involvement of the NLP4RE community for the extension of the study. \Cref{fig:process} visualizes the two-phase process.

\subsubsection{Study sample selection and preparation}
\label{sec:recovery:design:sample}

Prior studies~\cite{montgomery2022empirical,frattini2022live} provided an existing selection of primary studies relevant to the requirements quality literature. In particular, the subject of our recovery request is the artifacts from a set of primary studies that we used to construct an ontology of requirements quality factors~\cite{montgomery2022empirical}. To develop this ontology, we collected manuscripts reporting quality factors from an original set of publications reported in another secondary study~\cite{montgomery2022empirical}. Extracting data sets and implementations from such publications revealed the unfortunate state of artifact availability in the first place. Reusing this set of artifacts qualifies our sampling strategy as \textit{convenience sampling}~\cite{baltes2022sampling}.

In the initial ontology-creation study~\cite{frattini2022live}, we extracted only the name of the artifact, its availability, and---in case it was accessible---its source link. To enable recovery requests for each unavailable resource, we additionally extracted the following information for each artifact:

\begin{itemize}
    \item \textbf{Corresponding author}: each artifact was associated with a corresponding author responsible for its availability.
    \item \textbf{Mention}: each artifact was associated with its verbatim mention in the manuscript.
\end{itemize}

Additionally, we corrected information about one data set and three implementations that persisted in the previous study.\footnote{All corrections are documented in the replication package.} In three spreadsheets, we collected data about

\begin{itemize}
    \item (1) \textbf{authors} (n=35), specified by their name and email address,
    \item (2) \textbf{data sets} (n=57), and (3) \textbf{implementations} (n=36), specified by the publication which contains it, its verbatim mention, the corresponding author, and its current availability.
\end{itemize}

\subsubsection{Approaching the authors}
\label{sec:recovery:design:approaching}

We created a Python script that automatically assembles one email to each corresponding author. This email contained the following elements:

\begin{enumerate}
    \item Header: an explanation of our endeavor and a request to contribute to open science (or alternatively explain why recovery is impossible).
    \item Artifact list: a list of artifacts for which the corresponding author was responsible
    \item Instructions: brief instructions on how to properly disclose artifacts according to open science principles as well as the offer to assist them in the process
    \item Contact: a way to reach out to us
\end{enumerate}

We developed the instructions based on our collective knowledge regarding open science, relevant literature~\cite{rosen2005open}, and artifact evaluation guidelines of SE conferences.
These instructions are a condensed version of the management guideline presented later (see \Cref{sec:guide}) and are contained in our replication package.
The development of the instructions is described in more detail in \Cref{sec:guide:method}.

In the initial phase of the recovery attempt, we approached the authors in a first mail on the 30\textsuperscript{th} of November 2022, followed by a reminder on the 13\textsuperscript{th} of December, and a final reminder on the 11\textsuperscript{th} of January 2023. For authors who did not respond to our request until the final reminder, we additionally contacted their co-authors to increase the likelihood of response. We concluded the recovery process on the 8\textsuperscript{th} of February 2023, yielding a time frame of 70 days.

\subsubsection{Correspondence}
\label{sec:recovery:design:correspondence}

We kept close contact with the authors we approached by responding in a window of 24 hours within workdays. During this process, we clarified concerns and offered our help. We processed and recorded the information contained in the authors' answers in a spreadsheet file. We tracked the response status to our request in an additional column, denoting the request as either \textit{undeliverable}, \textit{unanswered}, \textit{answered}, or \textit{completed}. We labeled a recovery request as \textit{completed} once the corresponding author, for all their artifacts, either improved their availability or explained the inability to recover or disclose them.

Furthermore, we documented the dates of the first email sent, the first response received, and the completion of the request alongside the number of emails sent by the author in addition to the updated availability status of the artifacts or, eventually, the author's explanation for not taking the recommended actions. Two authors coded these explanations independently and came to an absolute agreement on the types of reasons for non-recovery. When the corresponding author's email address was no longer used, we reached out via personal contacts or social networks like Twitter and LinkedIn.

\subsubsection{Evaluation}
\label{sec:recovery:design:evaluation}

To evaluate the artifact recovery process, we generated statistics of the following data from the documentation in our tables. 

\begin{enumerate}
    \item Correspondence (i.e., author response time and frequency) to evaluate the effort of the recovery process.
    \item Recovery request success (i.e., change in artifact availability) to evaluate the success of the recovery process.
    \item Reason for non-recovery (i.e., authors' responses excusing the recovery) to evaluate the reasons inhibiting adherence to open science principles.
\end{enumerate}

We evaluated the data by generating descriptive statistics from our documentation.

\subsubsection{Dissemination and Crowd-Sourcing}
\label{sec:recovery:design:crowdsourcing}

The results of the first phase of the recovery request showed initial success but also room for improvement~\cite{frattini2023let} (more details in \Cref{sec:recovery:result}). The first author of this article presented these results at the 6\textsuperscript{th} Workshop on Natural Language Processing for Requirements Engineering\footnote{\url{https://nlp4re.github.io/2023/}} (NLP4RE) co-located with the 29th International Working Conference on Requirement Engineering: Foundation for Software Quality\footnote{\url{https://2023.refsq.org/}} (REFSQ). The visualization of the dire previous state of open science in the research field, but also the initial success of the artifact recovery, inspired the workshop attendees to contribute to the recovery initiative. Three hypotheses about the remaining room for improvement emerged:

\begin{enumerate}
    \item Corresponding authors might not have responded to the recovery request because its sender was unknown to them, and they were unable to verify the trustworthiness of the request.
    \item Corresponding authors might be more likely to react to an alternative email address than the corresponding email address printed in a published article. For example, some workshop attendees were aware of email addresses of unresponsive authors through which they could personally reach out.
    \item Artifacts lost to corresponding authors might have been acquired by members of the community at the time they were available.
\end{enumerate}

These three hypotheses invited crowd-sourcing the recovery initiative, as the more established members of the research community are likely to have a better connection to the corresponding authors of unavailable artifacts (addressing hypotheses 1 and 2), and they might have acquired resources and could still recover them when corresponding authors have already confirmed their status of unavailability (addressing hypothesis 3). To this end, we compiled two lists: one containing unresponsive corresponding authors and one containing artifacts that corresponding authors claimed to be lost. The lists were distributed to seven attendees of the NLP4RE workshop who expressed interest in contributing to the recovery initiative. All of those contributors are experienced and established scholars in the research domain.

The contributors partook in the artifact recovery initiative by providing the authors of this paper with additional information, establishing contacts, and recovering artifacts. Once a new contact was established, the artifact recovery request was handled as in the first phase of the initiative. Once a contributor retrieved an artifact, the owner of the artifact was determined and approached to request permission for archiving it via Zenodo.\footnote{\url{https://zenodo.org/}}

We initiated the second, crowd-sourced phase of the artifact recovery initiative on the 2\textsuperscript{nd} of June 2023, sent a reminder on the 19\textsuperscript{th} of June 2023, and concluded all requests on the 30\textsuperscript{th} of June 2023. We evaluated the recorded data similar to the first phase described in \Cref{sec:recovery:design:evaluation}. Note that we merely used the above-mentioned three hypotheses to design the second phase of the artifact recovery initiative and did not empirically evaluate them since we prioritized the recovery of research artifacts over investigating research community dynamics.
 
\subsection{Results}
\label{sec:recovery:result}

\subsubsection{Correspondence}
\label{sec:recovery:result:correspondence}

\begin{figure}
    \centering
    \includegraphics[width=0.9\textwidth]{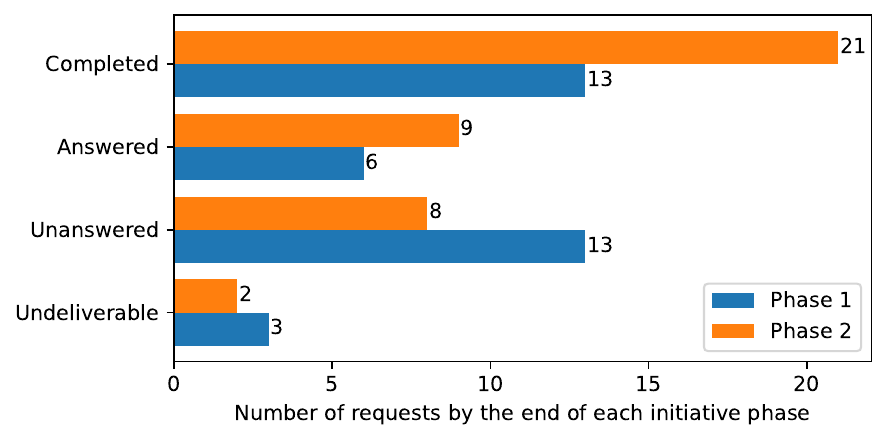}
    \caption{Status of correspondence}
    \label{fig:correspondence:status}
\end{figure}

\paragraph{Phase 1} Out of the 35 approached corresponding authors, 19 (54.3\%) answered the recovery request, and 13 (37.1\%) completed it. We could not reach three (8.6\%) authors despite searching for a valid contact. The distribution of correspondence status is visualized as the blue bars in \Cref{fig:correspondence:status}. 

\paragraph{Phase 2} During the second phase, \nauthorsnew\ additional corresponding authors were identified, as it became clear that \nauthorsnew\ artifacts were actually owned by other, previously not considered authors. Out of these \nauthors\ approached corresponding authors, \requestanswered\ (\requestansweredperc\%) answered the recovery request, and \requestcompleted\ (\requestcompletedperc\%) completed it. \requestundeliverable\ (\requestundeliverableperc\%) authors remained unreachable and \requestunanswered\ (\requestunansweredperc\%) requests remained unanswered. The distribution of correspondence status is visualized as the orange bars in \Cref{fig:correspondence:status}.

\begin{figure}
    \centering
    \includegraphics[width=0.9\textwidth]{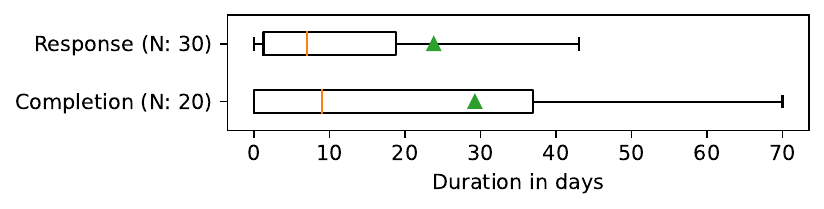}
    \caption{Distribution of time for correspondence in days (excluding outliers)}
    \label{fig:correspondence:time}
\end{figure}

\begin{figure}
    \centering
    \includegraphics[width=0.9\textwidth]{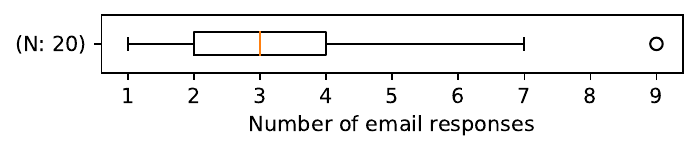}
    \caption{Distribution of frequency of correspondence in the number of emails}
    \label{fig:correspondence:freq}
\end{figure}

It took, on average, \daysresponse\ days for a corresponding author to reply to our request and \dayscomplete\ additional days to complete the request. On average, a request was resolved in an exchange of 3 emails with the corresponding author. The distributions of these statistics are visualized in \Cref{fig:correspondence:time} and \Cref{fig:correspondence:freq}, respectively.

\subsubsection{Artifact Recovery Success}
\label{sec:recovery:result:success}

\Cref{tab:recovery:summary} summarizes the total number of artifacts that were either recovered or where their unavailability was confirmed by the corresponding authors of primary studies over the two phases. 
An artifact counted as \textit{availability improved} if its availability at the end of the respective phase was higher than at the beginning of the study according to \Cref{tab:statuscodes}.
The availability of 10 data sets was improved through the two recovery phases.
Two of these 10 data sets were already available before but on a lower level of \Cref{tab:statuscodes}, and eight data sets were newly recovered.
This increases the availability of data sets from 12.3\% (7/57, 1 archived) to \datasetsavailableperc\% ((7+8)/57=\datasetsavailable/57, \datasetsarchived\ archived). 
Similarly, the availability of 8 implementations was improved.
Five of these were not available before. 
Additionally, one previously available implementation became unavailable during the study (see the explanation for this below), such that
the availability of implementations improved from 19.4\% (7/36, 0 archived) to \implementationsavailableperc\% ((7+5-1)/36=\implementationsavailable/36, \implementationsarchived\ archived). Authors further confirmed the unavailability of \datasetsclarified\ (\datasetsclarifiedperc\%) data sets and \implementationsclarified\ (\implementationsclarifiedperc\%) implementations and provided reasons for the inability to recover or disclose them.

\begin{table}[ht]
    \centering
    \footnotesize
    \caption{Total number of artifacts clarified after each of the two phases}
    \label{tab:recovery:summary}
    \begin{tabular}{l|rr|rr}
        \textbf{Availability} & \multicolumn{2}{c|}{\textbf{Phase 1}} & \multicolumn{2}{c}{\textbf{Phase 2}} \\
         & (D) & (I) & (D) & (I) \\ \hline
        Availability improved & 7 (12.3\%) & 7 (19.4\%) & \datasetsimproved\ (\datasetsimprovedperc\%) & \implementationsimproved\ (\implementationsimprovedperc\%) \\
        Unavailability confirmed & 21 (36.8\%) & 6 (16.7\%) & \datasetsclarified\ (\datasetsclarifiedperc\%) & \implementationsclarified\ (\implementationsclarifiedperc\%) \\ \hline
        \textbf{Total} & 28 (49.1\%) & 13 (36.1\%) & \datasetsaddressed\ (\datasetsaddressedperc\%) & \implementationsaddressed\ (\implementationsaddressedperc\%)
    \end{tabular}
\end{table}

\begin{figure}[ht]
    \centering
    \includegraphics[width=0.8\textwidth]{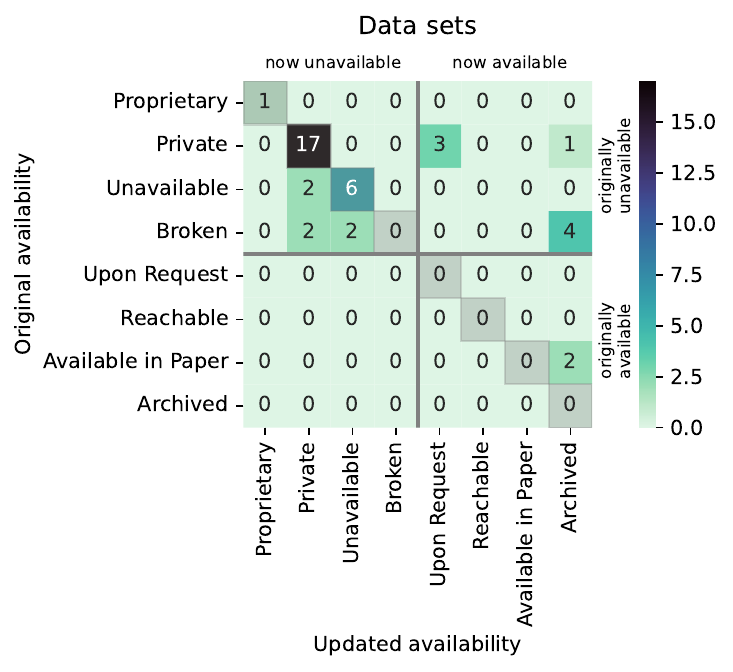}
    \caption{Change of availability in data sets}
    \label{fig:availability:datasets}
\end{figure}

\begin{figure}[ht]
    \centering
    \includegraphics[width=0.8\textwidth]{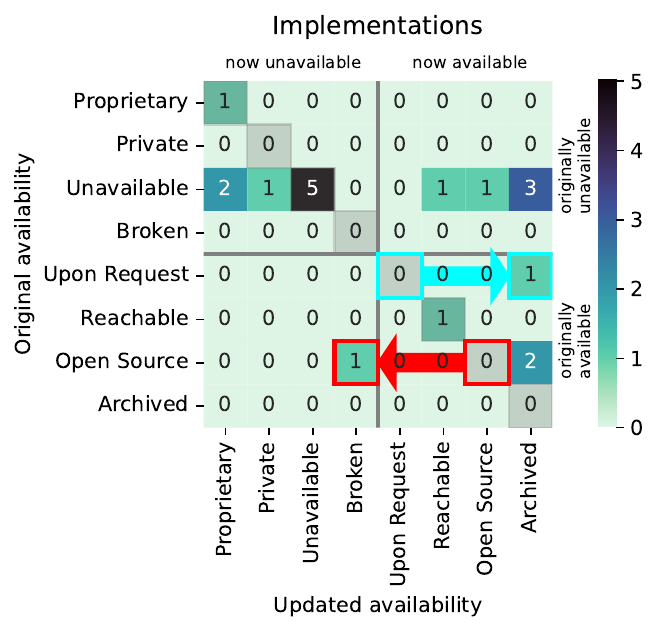}
    \caption{Change of availability in implementations}
    \label{fig:availability:approaches}
\end{figure}

\Cref{fig:availability:datasets,fig:availability:approaches} visualize the total success of the two-phase recovery initiative. The heatmap considers all artifacts (data sets in \Cref{fig:availability:datasets} and implementations in \Cref{fig:availability:approaches}) where the corresponding author completed the recovery request. The number in a cell represents the number of artifacts for which the original availability (on the y-axis) has been updated to the new availability (on the x-axis). The count of artifacts whose availability remained the same (e.g., because an author confirmed that the artifact could not be made more available) is reported on the diagonal (shaded gray). An improvement in the availability of an artifact contributes to cells to the right of the diagonal, a deterioration of the availability to the left.
Consequently, the ``Availability improved'' count in \Cref{tab:recovery:summary} is the sum of all cell values to the right of the diagonal, and the ``Availability confirmed'' count is the sum of all cell values on the diagonal and to the left of the diagonal.

For example, one implementation was previously available upon request~\cite{li2016engineering}. Now that the authors archived the implementation following open science principles,\footnote{Now publicly available at \url{https://doi.org/10.5281/zenodo.7484023}} the entry moved three cells to the right (see the cyan arrow in \Cref{fig:availability:approaches}).
On the other hand, another implementation called RETA~\cite{arora2015automated} was available at \url{http://sites.google.com/site/retanlp/} during the first phase of the study~\cite{frattini2023let}. While checking during the second phase, the URL did no longer resolve and the link became \textit{broken}. The entry, therefore, moved three cells to the left (see the red arrow in \Cref{fig:availability:approaches}). The authors explained the loss of the artifact with their change of affiliation, which caused the website not to be maintained anymore. The implementation could sadly not be recovered at the time.

The inability to recover or disclose artifacts was reported as follows: among \datasetsclarified\ unrecoverable data sets, \datasetslost\ were lost (i.e., the author could not find them anymore or the contact of whom the author assumed had the data was unreachable), and \datasetssensitive\ could not be disclosed due to sensitive contents. Among the \implementationsclarified\ unrecoverable implementations, \implementationsproprietary\ became proprietary, and \implementationslost\ were lost.

The recovered artifacts (i.e., all artifacts that counted towards ``Availability improved'' in \Cref{tab:recovery:summary}), their new location, and the original publication presenting them are listed in \Cref{tab:recovered:data} (data sets) and \Cref{tab:recovered:implementations} (implementations).

\begin{table}[h]
    \centering
    \footnotesize
    \caption{List of and reference to recovered data sets}
    \label{tab:recovered:data}
    \begin{tabularx}{\textwidth}{p{3.2cm}|Xl}
        \textbf{Artifact} & \textbf{Link} & \textbf{Ref} \\ \hline
        UIC EIRENE SRS & \url{https://doi.org/10.5281/zenodo.1414117} & \cite{ferrari2013using} \\
        HWS Documentation & \url{https://doi.org/10.5281/zenodo.80815230} & \cite{rago2014approach} \\
        3 user story SRS & (upon request) & \cite{lucassen2017improving} \\
        Real estate SRS & \url{https://doi.org/10.5281/zenodo.7499290} & \cite{femmer2014impact} \\
        IMAGS II SRS & \url{https://doi.org/10.5281/zenodo.7619051} & \cite{din2008requirements} \\
        11 SRS & \url{https://doi.org/10.5281/zenodo.80143477} & \cite{yang2012speculative} \\
        MadeByGraph SRS & \url{https://doi.org/10.5281/zenodo.7602827} & \cite{mokammel2018automatic} \\
        Helpdesk Support & \url{https://doi.org/10.5281/zenodo.8183338} & \cite{al2008identifying} \\
    \end{tabularx}
\end{table}

\begin{table}[h]
    \centering
    \footnotesize
    \caption{List of and reference to recovered implementations}
    \label{tab:recovered:implementations}
    \begin{tabularx}{\textwidth}{p{3.2cm}|Xl}
        \textbf{Artifact} & \textbf{Link} & \textbf{Ref} \\ \hline
        S-HTC &\url{https://doi.org/10.5281/zenodo.7584181} & \cite{ferrari2013using} \\ 
        CAR & \url{https://doi.org/10.5281/zenodo.7584193} & \cite{ferrari2014measuring} \\
        AQUSA & \url{https://doi.org/10.5281/zenodo.7573781} & \cite{lucassen2017improving} \\
        Bidirectional Chatbot Cordula & \url{https://git.uni-paderborn.de/jkers/sfb-b1-cordula-bidirectional} & \cite{baumer2018flexible} \\
        Desiree & \url{https://doi.org/10.5281/zenodo.7484023} & \cite{li2016engineering} \\
        ARBIUM & \url{https://doi.org/10.5281/zenodo.7528522} & \cite{el2010improving} \\
        Ambiguity detector & \url{https://doi.org/10.5281/zenodo.1476902} & \cite{ferrari2019nlp} \\
        Near-synonymy detector & \url{https://github.com/RELabUU/revv} & \cite{dalpiaz2018pinpointing} \\
    \end{tabularx}
\end{table}

\begin{highlightbox}{Answer to RQ1: Success of the Recovery Request}
    Approaching owners of unavailable artifacts with a request to recover them showed significant success in the requirements quality research community. 
    Crowd-sourcing the request in the research community further benefited the endeavor.
    The status of several unavailable research artifacts could be clarified this way, either by explaining their unavailability with valid reasons or by recovering the artifacts.
\end{highlightbox}

\subsection{Threats to Validity}
\label{sec:recovery:threats}

\revmeta{The answer to RQ1, as stated in \mbox{\Cref{sec:recovery:result}}, is subject to the following threats to validity, grouped via the categories introduced by Wohlin et al.~\mbox{\cite{wohlin2012experimentation}}.
The main type of threat affecting the validity of our claim is \textit{external validity}.
The generalizability of the claim is inhibited by the sample of primary studies involved in the recovery attempt.
The sample originates from previous secondary studies~\mbox{\cite{montgomery2022empirical,frattini2022live}}.
The representativeness of our sample is inherently limited by their cutoff date, the 27th of March 2020.
However, the original secondary study~\mbox{\cite{montgomery2022empirical}} uses a rigorous sampling strategy, which supports the reliability of the claim at least for the respective time frame.}

\revmeta{Additionally, threats to \textit{internal validity} may affect the conclusions of the study.
Confounding factors could have affected the result.
For example, we are unable to explain the different reasons that caused the non-responsiveness of some authors.
Possible factors include negligence, distrust, or oversight.
Given the setup of the study, we could not control these factors.
}

\section{Evaluation of Reasons for Artifact Unavailability}
\label{sec:eval}

To infer deeper insight into the factors influencing artifact (un-)availability in our sample, we analyzed the data from the two-phase recovery initiative. For our sample of research artifacts from the requirements quality research domain, we ask the following research questions:

\begin{itemize}
    \item RQ2: Which factors influence the availability of research artifacts?
    \item RQ3: Which factors influence the success of the recovery initiative?
\end{itemize}

\Cref{sec:eval:vars} states our hypotheses and summarizes the available variables, and \Cref{sec:eval:bda} documents the analysis procedure. \Cref{sec:eval:results} contains the results and \Cref{sec:eval:interpretation} interprets them, before \Cref{sec:eval:ttv} discusses threats to validity of this analysis.

\subsection{Hypotheses and Variables}
\label{sec:eval:vars}

The collected data allows us to infer insights about four aspects of artifact availability. Within our sample of observed artifacts, we investigate the factors that influence

\begin{enumerate}
    \item the \textbf{original availability} of artifacts (\textit{orig}), i.e., the inclination of authors to disclose artifacts upon publication of their article (regardless of the longevity of this artifact),
    \item the \textbf{persistence} of disclosed artifacts (\textit{per}), i.e., the longevity of an artifact,
    \item the \textbf{recoverability} of an unavailable artifact (\textit{recov}), i.e., the ability to make it available again, and
    \item the \textbf{updated availability} of all artifacts (\textit{avail}), i.e., the accessibility of artifacts after the recovery initiative.
\end{enumerate}

The original availability of artifacts is determined by whether a primary study contains a link to the disclosed artifact. 
In this case, it does not matter whether that link still resolves, as we assume that it at least resolved at the time of publication. The persistence is measured by whether a link to an artifact still resolves. 
The recoverability is measured by whether the recovery was successful or had to be excused by the corresponding author. 
The updated availability is measured via the availability status of the sampled artifacts after the recovery initiative.

The original availability and persistence of research artifacts represent the main variables of interest that motivated the recovery initiative in the first place.
Investigating hypotheses involving these two variables contributes insights into the adherence to open science principles and the artifact availability in requirements quality research.
The recoverability and updated availability of research artifacts represent variables describing the outcome of the recovery initiative itself.
Investigating hypotheses involving these two variables contributes insights about executing artifact recovery initiatives and helps decide whether the effort to implement such an initiative is worthwhile.

Given the collected data, the impact of the following independent variables on the presented dependent variables can be investigated:

\begin{enumerate}
    \item \textbf{Recency} ($rec$): Relative age of the publication presenting the artifact
    \item \textbf{Type} ($type$): Whether the artifact is a data set or implementation
    \item \textbf{Hosting} ($host$): Whether an artifact was hosted using a public (e.g., Zenodo, Github, Sourceforge) or private (e.g., institutional or personal websites) service
\end{enumerate}

The category \textit{public} of the independent variable \textbf{hosting} could further differentiate \textit{archival}, i.e., whether the public host is committed to a long-term retention policy, similar to what Winter et al.~\cite{winter2022retrospective} investigated.
Due to the lack of data points, we were unable to consider this differentiation.

Further independent variables could causally impact the dependent variables, like the artifact policy of venues at the time of publication or a corresponding author's previous knowledge of open science principles and practices. We did not collect additional data beyond the documentation of the artifact recovery initiative and confined our inference to the available variables. This limits our causal inference---which we further discuss in the threats to validity in \Cref{sec:eval:ttv}---but still allows limited reasoning within an explicitly delineated space of variables.

The variables and their data types are summarized in \Cref{tab:vars}. For the three of the total four dependent variables \textit{original availability} ($orig$), \textit{recoverability} ($recov$), and \textit{updated availability} ($avail$), we investigate the impact of the two independent variables \textit{recency} ($rec$) and \textit{type} ($type$). Consequently, we formulate the following six hypotheses: ``The \{\textit{recency} of a publication, \textit{type} of an artifact\} has no effect on the \{\textit{original availability}, \textit{recoverability}, \textit{updated availability}\} of an artifact'' ($h_{dep \in \{orig, recov, avail\}}^{ind \in \{rec, type\}}$). For the fourth dependent variable \textit{persistence} ($per$), we investigate the impact of the independent variables \textit{recency} ($rec$) and \textit{hosting} ($host$) in the scope of the following hypothesis: ``The \{\textit{recency} of a publication, \textit{hosting} of an artifact\} has no effect on the \textit{persistence} of the artifact'' ($h_{per}^{ind \in \{rec, host\}}$).
This results in eight hypotheses to be tested in this evaluation.
We index hypotheses based on the combination of independent and dependent variables under investigation, e.g., $h_{orig}^{rec}$: ``The \textit{recency} of a publication has no effect on the \textit{original availability} of an artifact.''
For conciseness, we keep the rest of the hypotheses in their modular format.
They are spelled out in the respective analysis files contained in our replication package.

\begin{table}[hbt!]
    \centering
    \footnotesize
    \caption{Independent (ind) and dependent (dep) variables for empirical analysis}
    \label{tab:vars}
    \begin{tabular}{p{2.2cm}|l|c|p{4cm}|p{2cm}}
        \textbf{Variable} & \textbf{Name} & \textbf{Type} & \textbf{Description} & \textbf{Range}\\ \hline
        Recency & recen & ind & Relative age of an artifact & [0, 1] \\
        Type & type & ind & Type of the artifact & \{dataset, implementation\} \\
        Hosting & host & ind & Type of artifact hosting &  [private, public] \\
        \hline
        Original Availability & orig & dep & The availability of an artifact before the recovery initiative & \{available, unavailable\} \\
        Persistence & per & dep &  Whether a once disclosed artifact was available before the recovery initiative &  \{persistent, non-persistent\} \\
        Recoverability & recov & dep & Whether the availability of an artifact could be improved & \{recoverable, unrecoverable\} \\
        Updated Availability & avail & dep & The availability of an artifact after the recovery initiative & \{available, unavailable\} \\
    \end{tabular}
\end{table}

\subsection{Bayesian Data Analysis}
\label{sec:eval:bda}

We conducted a Bayesian data analysis (BDA) according to Pearl's framework for causal inference~\cite{glymour2016causal}. Furia et al. have popularized the use of BDA in software engineering for causal inference since it outperforms common frequentist approaches in terms of reliability and level of detail of the inference. Most importantly, BDA models unknown parameters as probability distributions instead of fixed values, allowing for more sophisticated insights than point-wise comparisons of frequentist statistical methods~\cite{furia2019bayesian}.
For brevity, we only report the most important elements of the analysis in this manuscript.
We refer the reader interested in a gentler introduction to the topic to appropriate literature on frameworks for causal inference~\cite{glymour2016causal,siebert2023applications}, textbooks on BDA~\cite{mcelreath2020statistical}, exemplary applications of BDA in SE research~\cite{furia2019bayesian,furia2022applying}, and our replication package mentioned in the introduction.

\begin{figure}
    \centering
    \includegraphics[width=0.6\textwidth]{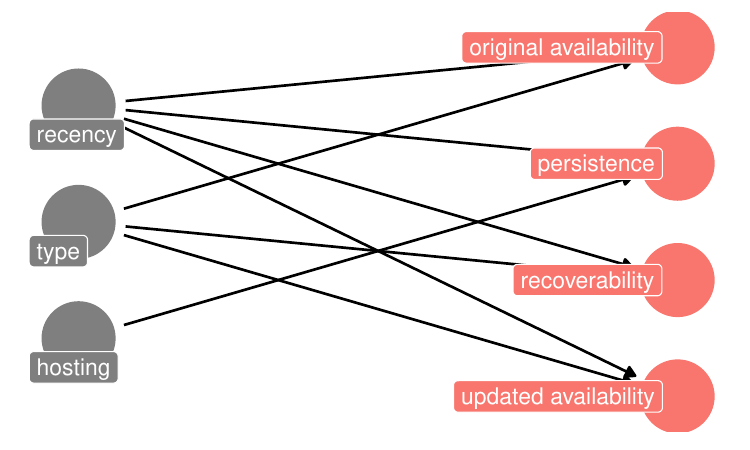}
    \caption{Directed, acyclic graph of causal assumptions between independent (grey) and dependent (red) variables}
    \label{fig:dag}
\end{figure}

We implemented the BDA according to an established Bayesian workflow in three major steps~\cite{siebert2023applications}: modeling, identification, and estimation. In the first \textit{modeling} step, we formalized our causal assumptions as a directed, acyclic graph (DAG) as visualized in \Cref{fig:dag}. In the DAG, nodes represent the variables of interest, and directed edges represent assumed causal relationships between these nodes. Based on the DAG, variables to include and exclude can be determined via selection criteria~\cite{mcelreath2020statistical} in the \textit{identification} phase. Because our DAG does not involve any variables that influence both the independent and the dependent variables, all variables are eligible to be included in the next phase. 

In the final \textit{estimation} phase, we trained Bayesian models following the Bayesian workflow by Gelman et al.~\cite{gelman2020bayesian} and using the R library \texttt{brms}~\cite{burkner2017brms}. Each dependent variable was modeled in relationship to its independent variables, which allows to quantify the impact of each independent variable on the dependent variable after training.

We selected an appropriate likelihood distribution for each dependent variable according to the maximum entropy criterion~\cite{jaynes2003probability}. Since all dependent variables are categorical with two categories, a single-trial Binomial or Bernoulli distribution is appropriate~\cite{mcelreath2020statistical}. We opt for the latter due to simplicity. Next, we selected uninformative priors, i.e., parameter distributions that represent prior beliefs but are unspecific enough for the model to update these beliefs according to the data~\cite{gelman2020bayesian}. We confirmed the appropriateness of these priors through graphical prior predictive checks~\cite{wesner2021choosing}.

We then trained all models on the collected data and assessed the appropriateness of the trained models both through graphical posterior predictive checks~\cite{gelman2020bayesian} and assessment of the chain mixing property $\hat{R} < 1.01$~\cite{vehtari2021rank}. Then, we evaluated our models by plotting the conditional effects of the independent variables. 
The conditional effects represent the impact of the independent variables on a dependent variable as well as their interaction as learned by the model via Bayesian inference. It, therefore, quantifies the strength and direction of the relationship between independent and dependent variables. 

Since these conditional effects only visualize the uncertainty of a single variable, we additionally evaluate the model by sampling from the full posterior distribution of the learned model parameters~\cite{gelman2020bayesian}. For each dependent variable, we compare the samples from the posterior when fixing each independent variable at the extreme values of its spectrum (0 and 1 in the case of \textit{recency}, and both categorical values in the case of \textit{type} and \textit{hosting}). By drawing 60.000 random samples from the posterior distribution once for each of the two extreme values, the model allows to quantify the impact of each independent variable in terms of percentages.

For more in-depth explanations of Bayesian data analysis, we refer the interested reader to established literature~\cite{mcelreath2020statistical,gelman2020bayesian,furia2019bayesian,glymour2016causal} and our replication package containing detailed documentation of the analysis.

\subsection{Results}
\label{sec:eval:results}

The following sections contain the evaluation of the hypotheses ($h_{orig}$ in \Cref{sec:eval:results:orig}, $h_{per}$ in \Cref{sec:eval:results:per}, $h_{recov}$ in \Cref{sec:eval:results:recov}, and $h_{avail}$ in \Cref{sec:eval:results:ava}). For brevity, we removed figures not contributing significantly to the results. All visualizations and evaluations are, however, available in our replication package.

\subsubsection{Factors influencing original Availability}
\label{sec:eval:results:orig}

\begin{figure}[htb]
    \centering
    \begin{subfigure}[b]{0.49\textwidth}
        \centering
        \includegraphics[width=\textwidth]{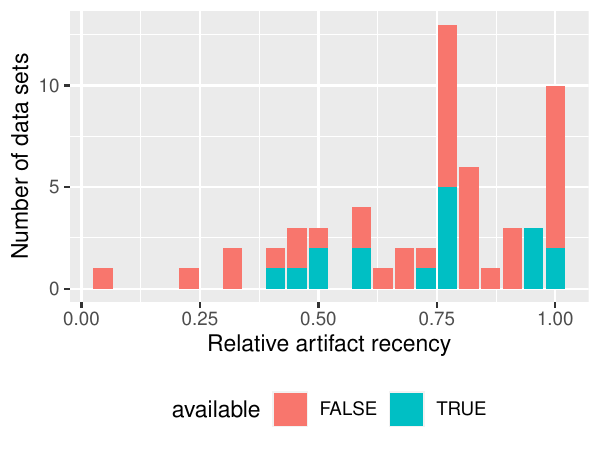}
        \caption{Distribution of data set availability}
        \label{fig:original:data}
    \end{subfigure}
    \hfill
    \begin{subfigure}[b]{0.49\textwidth}
        \centering
        \includegraphics[width=\textwidth]{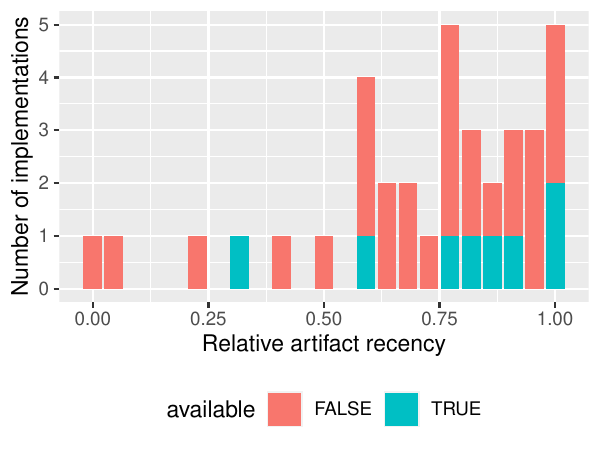}
        \caption{Distribution of implementation availability}
        \label{fig:original:implementation}
    \end{subfigure}
    \caption{Visualization of original availability}
    \label{fig:original}
\end{figure}

\Cref{fig:original} visualizes the distribution of the original availability of data sets and implementations over the time span of the articles. The value $recency=0$ represents the oldest data set (1998, a requirements specification presented by Romano et al.~\cite{romano1998tbrim}) and implementation (1997, the ARM tool by Wilson et al.~\cite{wilson1997automated}) respectively, $recency=1$ represents the most recent data set (2019, several requirements specifications by Wang et al.~\cite{wang2019paser}) and implementation (2019, the PASER tool by Wang et al.~\cite{wang2019paser}). An artifact was considered available when its availability status code (see \Cref{tab:statuscodes}) was \textit{broken} or better, as we assume that now-broken links worked upon the initial publication of the primary study. 


\begin{figure}
    \centering
    \includegraphics[width=0.6\textwidth]{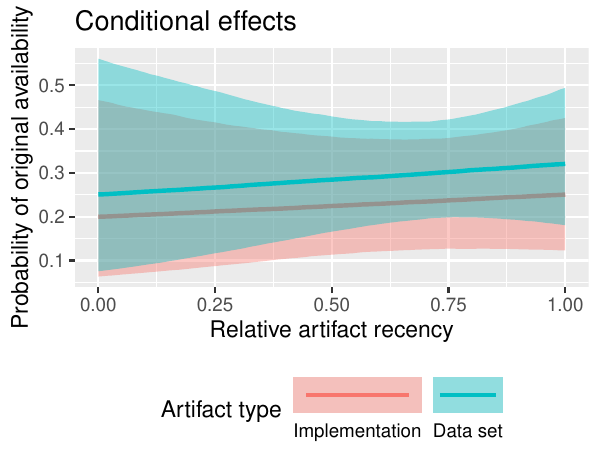}
    \caption{Conditional effect between recency and artifact type on original availability}
    \label{fig:original:conditional}
\end{figure}

\Cref{fig:original:conditional} visualizes the conditional effect of both recency and the artifact type on the original availability of artifacts as picked up by the trained model. The plot shows that data sets are, in general, more likely to be available upon publication of a primary study and corroborates the impression that the original availability improved over recent years, i.e., authors were more inclined to disclose their artifacts upon publication. 

These inferences are further strengthened by the random samples from the posterior distribution: while data sets are---on average and taking into account all uncertainty of the Bayesian model---made available in around 29\%, implementations are only available in 23\% of the time. The original availability of artifacts, in general, is 24.6\% for the oldest and 29.2\% for the most recent publications, representing an average increase of original availability of around 5\%.

\subsubsection{Factors influencing Artifact Persistence}
\label{sec:eval:results:per}

\begin{figure}
    \centering
    \includegraphics[width=0.6\textwidth]{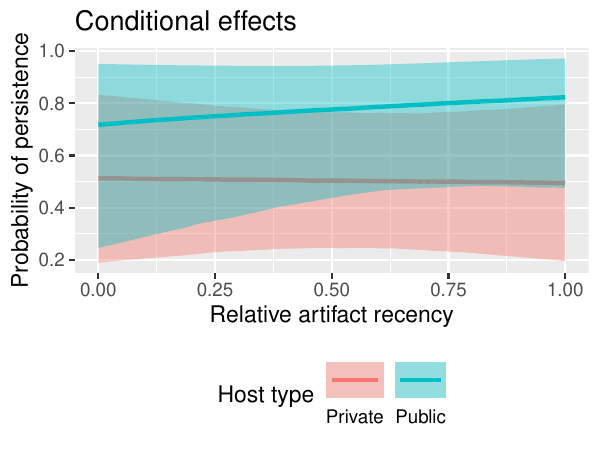}
    \caption{Conditional effect between recency and hosting type on artifact persistence}
    \label{fig:persistence:conditional}
\end{figure}

The raw data already shows that all cases of artifacts not persisting occurred using private hosting services: all three non-persistent artifacts were hosted on private services.
The sample of persistent artifacts is both privately and publicly hosted. 
All of the five artifacts hosted on public services had persisted.
The conditional effect in \Cref{fig:persistence:conditional} shows a strongly positive influence of public hosting services on the probability of artifact persistence, i.e., publicly hosted artifacts are much more likely to persist than privately hosted artifacts. According to the sampling from the posterior, privately hosted artifacts persist on average in 59\% of all cases, while publicly hosted artifacts persist in 76\%. Additionally, recency has a positive influence on persistence when using a public hosting service, while the impact is negligible for private hosting.

\begin{highlightbox}{Answer to RQ2: Factors influencing Artifact Availability}
    In the requirements quality research domain, more recent research articles are more likely to disclose research artifacts, and data sets are more likely to be disclosed than implementations.
    The use of a public hosting service has a positive impact on the persistence of these artifacts.
\end{highlightbox}

\subsubsection{Factors influencing Artifact Recoverability}
\label{sec:eval:results:recov}

\begin{figure}[htb]
    \centering
    \begin{subfigure}[b]{0.49\textwidth}
        \centering
        \includegraphics[width=\textwidth]{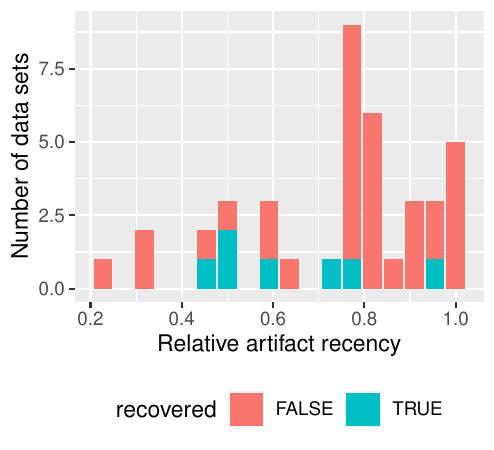}
        \caption{Distribution of data set recoverability}
        \label{fig:recoverability:data}
    \end{subfigure}
    \hfill
    \begin{subfigure}[b]{0.49\textwidth}
        \centering
        \includegraphics[width=\textwidth]{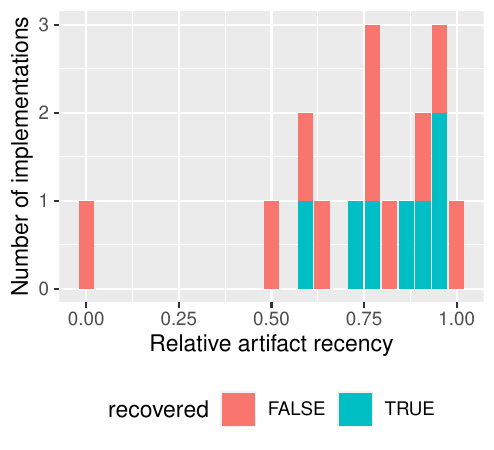}
        \caption{Distribution of implementation recoverability}
        \label{fig:recoverability:implementation}
    \end{subfigure}
    \caption{Visualization of Recoverability}
    \label{fig:recoverability}
\end{figure}

\begin{figure}
    \centering
    \includegraphics[width=0.6\textwidth]{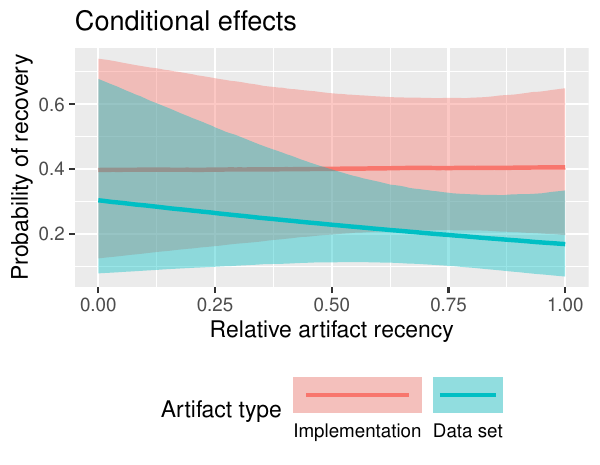}
    \caption{Conditional effect of recency and artifact type on recoverability}
    \label{fig:recoverability:conditional}
\end{figure}

\Cref{fig:recoverability} visualizes the distribution of recovered and non-recovered artifacts. The figure shows that the majority of artifacts addressed by their respective corresponding author during the artifact recovery initiative remained unrecoverable. The conditional effects in \Cref{fig:recoverability:conditional} show that implementations were, in general, more likely to be recovered. According to the posterior samples, implementations have a probability of recovery of 38\% while data sets only 23\%. Recency shows no significant influence on the recoverability of implementations but a negative influence on the recoverability of data sets. This means that more recent data sets were less likely to be recovered.

\subsubsection{Factors influencing overall Availability}
\label{sec:eval:results:ava}

\begin{figure}[htb]
    \centering
    \begin{subfigure}[b]{0.49\textwidth}
        \centering
        \includegraphics[width=\textwidth]{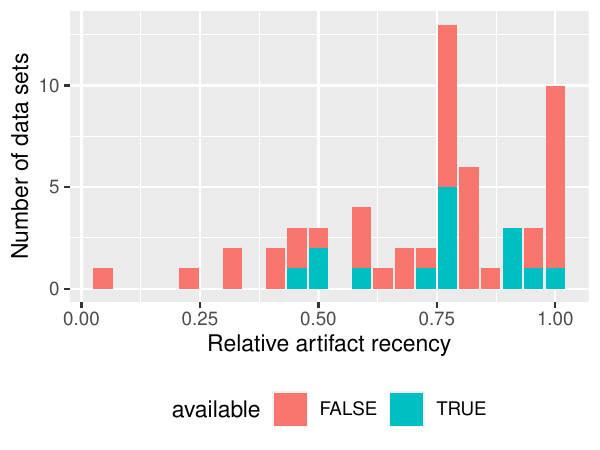}
        \caption{Distribution of data set availability}
        \label{fig:updated:data}
    \end{subfigure}
    \hfill
    \begin{subfigure}[b]{0.49\textwidth}
        \centering
        \includegraphics[width=\textwidth]{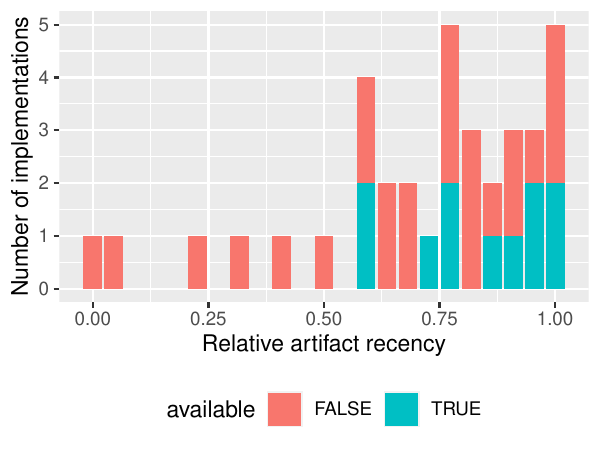}
        \caption{Distribution of implementation availability}
        \label{fig:updated:implementation}
    \end{subfigure}
    \caption{Visualization of updated availability}
    \label{fig:updated}
\end{figure}

\begin{figure}
    \centering
    \includegraphics[width=0.6\textwidth]{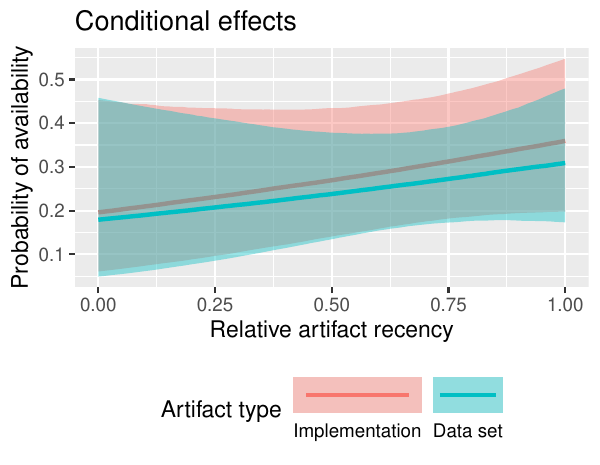}
    \caption{Conditional effect of recency and artifact type on overall availability}
    \label{fig:updated:conditional}
\end{figure}

\Cref{fig:updated} visualizes the distribution of the updated availability of artifacts in the sample, i.e., the overall availability after the artifact recovery initiative. The data shows and the conditional effects in \Cref{fig:updated:conditional} confirm that recency has a positive impact on overall artifact availability, i.e., more recent artifacts are more likely to be available. According to the sampling from the posterior, the least recent artifacts are only available with a probability of 21\% while the most recent with 33\%. In our sample, the model perceived the difference in updated availability between implementations and data sets as negligible.

\begin{highlightbox}{Answer to RQ3: Factors influencing Artifact Recovery}
    In the requirements quality research domain, implementations were more likely to be recovered than data sets.
    The recoverability of data sets decreased the more recent the publication in which they are contained was.
    After the two phases of the recovery initiative, data sets were equally likely to be available, and more recent artifacts were, overall, more likely to be available.
\end{highlightbox}

\subsection{Interpretation}
\label{sec:eval:interpretation}

The data analysis on our sample of 94 artifacts allows the following inferences: the recency of artifacts benefits the original ($h_{orig}^{rec}$) and updated availability ($h_{avail}^{rec}$) while showing a negative effect on artifact recoverability ($h_{recov}^{rec}$) for data sets. 
A follow-up hypothesis is that this is due to the increased use of sensitive, company-owned data in recent publications, which does not allow the recovery of unavailable artifacts.

Data sets are overall more likely to be disclosed upon publication of a study ($h_{orig}^{type}$) but less likely to be recovered ($h_{recov}^{type}$) if they were unavailable. 
This raises the following concern: implementations produced by the authors of a study are likely owned by these authors, in contrast to the (potentially private, company-owned) data to evaluate an implementation. 
The lack of available implementations despite a high likelihood of them being owned by the corresponding authors constitutes a clear opportunity for improvement in requirements quality research.

Finally, the use of a public hosting service strongly benefits the persistence of artifacts ($h_{per}^{host}$), which provides support for the advocacy of preferring dedicated artifact hosting services over institutional or private services.
As mentioned in \Cref{sec:eval:vars}, we cannot make a statement about the impact of long-term retention policies of public hosting services.
However, recent research from Winter et al.~\cite{winter2022retrospective} suggests that hosts committed to a long-term retention policy further improve the artifact availability.

The data evaluation shows that---despite the moderate success of the artifact recovery initiative---artifact availability remains improvable in the requirements quality research domain. 
Specifically, the strong positive effect of public hosting services on artifact persistence motivates greater effort in promoting adherence to open science principles and the adoption of open science tools. 
We address this need with our artifact management guideline in \Cref{sec:guide}.

\subsection{Threats to Validity}
\label{sec:eval:ttv}

The main threat to the validity of our empirical evaluation classifies as a threat to \textit{internal validity} according to the guideline by Wohlin et al.~\cite{wohlin2012experimentation}, i.e., the causal link between independent and dependent variables. We selected the independent variables for the empirical evaluation based on their availability as a result of the recovery initiative. Consequently, the dependent variables may be influenced by other independent variables that were not considered in the evaluation, for example, an individual researcher's knowledge of open science principles. We address this threat by adhering to the modeling step of our analysis framework~\cite{glymour2016causal,siebert2023applications}, which makes our causal assumptions and, therefore, the considered variables explicit. The DAG allows scrutinizing and extending our causal assumptions by including additional variables in future research~\cite{furia2019bayesian}.

\section{Open Science Artifact Management Guideline}
\label{sec:guide}

The lack of adherence to open science principles in the requirements quality research domain---especially despite the evident benefit of open science platforms like Zenodo~\cite{winter2022retrospective}---constitutes room for proactive improvement.
Especially the unavailability of software artifacts, which are mostly produced and owned by authors of a publication, is unfortunate in requirements quality research and undermines a core deliverable of this research area~\cite{krishnamurthi2015real}.

Since open science first reached the SE community roughly 10 years ago (2011/2013)~\cite{hermann2020community,krishnamurthi2015real}, the understanding and use of open science principles have evolved and grown.
This expansion necessitates a consolidation and communication of recommended practices in a simple, concise, and easy-to-read way.
For this, we have created the Artifact Management Guideline~\cite{montgomery2023artifact} presented in this section. \Cref{sec:guide:method} explains how the guideline was derived and \Cref{sec:guide:format} its format. \Cref{sec:guide:main} summarizes the guideline~\cite{montgomery2023artifact} and \Cref{sec:guide:vision} outlines its long-term vision. 
The guideline is archived at \url{https://doi.org/10.5281/zenodo.8134403}, and a collaborative version is accessible at \url{https://docs.google.com/document/d/1gIg3g-_zxCeiw2IJkBGbiGI9-3HeQU5FR63yAv3PhiM}.

\subsection{Method}
\label{sec:guide:method}

To derive this guideline, the authors combined their collective knowledge regarding open science~\cite{mendez2020open,frattini2023let} with a review of recent guidelines for artifact evaluation tracks (AET) at premiere SE research conferences as well as further material regarding open science in software engineering.
Most software engineering conferences have an AET~\cite{hermann2020community}, organizations such as ACM\footnote{\url{https://www.acm.org/publications/openaccess}} and IEEE\footnote{\url{https://open.ieee.org/about/}} have official open access policies, secondary-articles are being published regarding artifact availability~\cite{winter2022retrospective,frattini2023let}, meta-articles are being published with detailed descriptions and instructions on open science~\cite{mendez2019artifacts,mendez2020open}, and new ideas about open science, artifacts, and reuse have begun to arise~\cite{baldassarre2023re}.

Using the work of Hermann et al.~\cite{hermann2020community} as a starting point, we reviewed the AET guidelines of the International Conference on Software Engineering (ICSE), the International Conference on Requirements Engineering (RE), and the Joint European Software Engineering Conference and Symposium on the Foundations of Software Engineering (ESEC/FSE).
We utilized the last two years (2021/2022) of both their AET guidelines and their open science policy where available.
The work of Mendez et al.~\cite{mendez2019artifacts,mendez2020open} and Hermann et al.~\cite{hermann2020community} guided the conceptualization and framing of some of the sections.

\subsection{Chosen Format}
\label{sec:guide:format}

The purpose of this guideline is to deliver information in a simple, concise, comprehensible, and collaborative manner while following a collegial tone directed at the reader. 
In that sense, we opted for a more pragmatic way and writing style for the guideline and, in consequence, for subsequent sections of this manuscript.
We have chosen to create the guideline using GoogleDocs\footnote{\url{https://www.google.com/intl/en/docs/about/}} and archive major increments via Zenodo (see~\cite{montgomery2023artifact}). 
The document is set to receive comments, of which its maintainers will be notified and respond to within the document to foster interaction and collaboration.

\subsection{The Guideline}
\label{sec:guide:main}

The objective of the guideline is to advise scientists on how to collect, document, license, archive, and share artifacts.
``Artifacts'' includes---but is not limited to---scientific protocols, raw and derived data (text, CSV, etc.), scripts, figures, tables, and software.
The greater mission is to improve scientific rigor, encourage scientific collaboration, and increase the rate of scientific progression.

This guideline provides pragmatic support for the following five aspects of artifact management: collecting, documenting, licensing, archiving, and sharing.
\revmeta{\mbox{\Cref{tab:guideline}} summarizes the content of this guideline.}

\begin{table}
    \centering
    \caption{Guideline Overview}
    \label{tab:guideline}
    \begin{tabularx}{\textwidth}{llX}
        \toprule
        \textbf{Section} & \textbf{Aspect} & \textbf{Content} \\ \midrule
        \Cref{sec:guide:collect} & Collect & Gathering all data relevant to the scientific work, including: \\
        & $\bullet$ Open data & Raw and derived data, scientific protocols, tables, and extended findings \\
        & $\bullet$ Open material & Software tools, data collection, transformation, and analysis scripts \\
        & $\bullet$ Open access & Permanent, accessible, and unique identification of the manuscript and data \\ \hline
        \Cref{sec:guide:document} & Document & Creating a complete and coherent description of the artifact that allows installing, using, and evolving it \\ \hline
        \Cref{sec:guide:license} & License & Specifying the conditions under which the artifact and its constituents can be used \\ \hline
        \Cref{sec:guide:archive} & Archive & Hosting the artifact in a permanent and accessible way with a unique identifier (DOI) \\ \hline
        \Cref{sec:guide:share} & Share & Disseminating the artifact to invite the scientific community to use and evolve it, therefore contributing value to the community \\
        \bottomrule
    \end{tabularx}
\end{table}


\subsubsection{Collect}
\label{sec:guide:collect}

Scientific work is inherently exploratory and iterative, the result of which consists of drafted, incomplete, and often misplaced artifacts.
This necessitates finding, improving, and reviewing the artifacts associated with a scientific article.
The subject of this phase is the artifacts associated with open data, open material (inc. open source), and open access.
The collected artifacts should be placed in one folder and organized in a logical manner, such as methodological phases.

\paragraph{Open data}

Open data covers all data that contributed to the scientific claims made in an article, as future researchers need the exact data used in your scientific work to replicate, verify, and improve on scientific work.
This includes raw data, derived data, scientific protocols, but also figures, tables, and extended findings.

Raw data is used to generate or support claims in scientific work. 
This data tends to be untouched by the analysis, sometimes even before cleaning (since data cleaning may have introduced bias).
Derived data is created as a result of scientific analysis (automatically or manually), like models trained through machine learning algorithms and data created as a result of qualitative methods, such as coding tables, schemata, etc.
Scientific Protocols pertaining to the planning, execution, and adjustment of scientific work. 
This includes logs of decisions taken by participating scientists, protocols given to study participants, rationale for change requests, discussion notes, etc.
Figures and tables used to visualize results in the manuscript are as relevant to include as the code to generate them.
Furthermore, extended findings that did not fit into the manuscript fall into this category.

\paragraph{Open material}

Open material (including \textbf{open source}) covers all material that contributed to the scientific claims made in an article. 
Future researchers need these algorithms to replicate, verify, and improve on published work. 
This includes the following data collection scripts, data transformation scripts, analysis scripts, and software tools.
Data collection scripts are scripts used to collect research data, e.g., a custom web scraper, a script to iteratively access an API, or an HTML-to-SQL data-writing script.
Data transformation scripts are scripts used to transform data in unique ways, e.g., static analysis, machine learning, image recognition, or generative AI.
Analysis scripts are scripts used to analyze the final output data and potentially produce the published results. This includes scripts that generate figures or tables.
Software tools are used in the research, e.g., a custom survey tool, a new IDE, a Jira or GitLab plugin, etc.

\paragraph{Open access}
 
Finally, open access ensures that future users can find and access the associated article by including a permanent DOI link to an open-access article in the README, as well as the permanent DOI to the published version of the artifact. 
\Cref{sec:guide:document} contains details on how to document and \Cref{sec:guide:license} on open access licensing.

\subsubsection{Document}
\label{sec:guide:document}

Artifacts need to be documented such that they are approachable to someone unfamiliar with the employed workflow, development style, and organizational mindset.
At a minimum, this requires a \texttt{README.md} file in the top-level artifact folder containing the following sections:

\begin{itemize}
    \item \textbf{Summary of artifacts}: a concise description of the motivation and purpose of the artifact.
    \item \textbf{Author and article details}: a list of involved authors, including contact information such as emails and how to cite the work. This information may be subject to change as the proper citation string or the DOI of the article may not be known while preparing the artifact for submission. Provide as much information as possible at the current point in time and update the \texttt{README.md} once the information becomes available.
    \item \textbf{Description of artifacts}: an explanation of the folders and files, including what was not included (and why).
    \item \textbf{Licenses}: the chosen licenses for the artifacts (see \Cref{sec:guide:license} for more details).
\end{itemize}

If using the artifact requires more than opening PDFs, CSVs, etc., then an \texttt{INSTALL.md} file (also located in the top-level artifact folder) becomes necessary.
It should include the following sections:

\begin{itemize}
    \item \textbf{System requirements}: a generalization of the environment and programs necessary to execute any software or scripts. 
    \item \textbf{Installation instructions}: an instruction on how to execute the software or scripts in question. If possible, a virtualized setup (e.g., via Docker or a virtual environment) shall be provided. 
    \item \textbf{Steps to reproduce}: commands on how to reproduce the data, figures, tables, or results presented in the article. If the artifact is simple enough to not necessitate an \texttt{INSTALL.md} file, this section can be moved to the \texttt{README.md} file. 
\end{itemize}

Additional visualizations of the artifacts, e.g., a UML diagram of a software system, aid their accessibility and understandability~\cite{abrahao2012assessing,ducasse2005class,scanniello2018software}.

\subsubsection{License}
\label{sec:guide:license}

An essential part of sharing artifacts is an explicit statement about their (re)use as determined by licenses.
Open data requires a license attached to explicitly describe how third parties can use the data.
The Creative Commons (CC) licenses\footnote{\url{https://creativecommons.org/about/cclicenses/}} are often employed licenses for open data and offer a variety of regulations determining what third parties are allowed to do with the data.


The license applied to open material takes on the form of an open-source license, which applies specifically to source code.
Several online resources assist the selection of an open source license.\footnote{E.g. \url{https://choosealicense.com} and \url{https://opensource.org/licenses}}

The copyright license applied to an article itself is already decided through the copyright agreement with the publisher (IEEE, ACM, Elsevier, etc.). 
Understanding an author's rights over their work, particularly across the different versions of your article (``author-submitted article'', ``accepted article'', ``final published version'', etc.), is essential in ensuring open access to research articles. 
Tools like Sherpa ROMEO (now just ``Sherpa''),\footnote{\url{https://beta.sherpa.ac.uk/}} assist in checking compliance with publisher copyright models.


Once appropriate licenses are determined for all artifacts of an article---including the identification of licenses of not self-owned by reused, external artifacts---a section in the \texttt{README.md} document should state and explain the chosen licenses as discussed in \Cref{sec:guide:document}.
An additional and encouraged norm is to obtain the licenses as text files and place them in the artifact folder.


\subsubsection{Archive}
\label{sec:guide:archive}

A critical step to artifact availability is to upload it to a publicly available archival website. 
A hosting website should be selected when it meets all of the following three criteria:

\begin{enumerate}
    \item \textbf{Hosted online for public access}: Artifacts are hosted online for anyone to access via the internet. Additionally, there is no need for registration to access the artifacts.\footnote{``Free registration'' is not acceptable, as it is an additional barrier to obtaining the artifacts, and ``free'' often involves hidden clauses.}
    \item \textbf{Dedicated DOIs and immutable data}: DOIs are automatically created for artifacts, and both the DOIs and data they point to are immutable. Artifacts can be updated, but each version must be maintained with its own DOI.
    \item \textbf{Long-term maintainability}: The organization hosting the URL has committed to a long-term retention policy, i.e., it plans to maintain the artifact for the foreseeable future. For example, Zenodo states in its policies that ``Items will be retained for the lifetime of the repository''~\cite{zenodo2023policy} which covers ``the next 20 years at least''~\cite{zenodo2023policy}.
\end{enumerate}

In principle, any organization that fulfills the above requirements can be used for archiving artifacts.
In reality, there are currently only a few known organizations that satisfy these requirements: ArXiv for articles, and Zenodo and FigShare\footnote{Note that FigShare is a ``for profit'' commercial organization, which may affect long-term maintainability.} for all types of artifacts.

The following notable services do usually not satisfy the above requirements despite their recurring use.
Institutional websites,\footnote{Institutional websites can conform to the three open science requirements for archival as described above, in which case they would qualify for archival. However, this is usually not the case.} employee web pages and research group websites usually do not satisfy criterion 2 and 3, since institutions update their websites over time and do not maintain access to resources and URLs as supported by previous research by Winter et al.~\cite{winter2022retrospective}. 
Employee web pages are taken offline when the employee leaves. 
Similar problems exist for research group websites.
Cloud storage providers, such as Dropbox, Google Drive, OneDrive, and iCloud, are solutions for backing up and/or syncing data, not archiving data. 
For example, individuals can change the data and URLs at any time.
GitHub, GitLab, and other Social Git/Code Platforms offer features for social product development, which conflict with the requirements for open science artifact archival.
For example, they allow repositories to be deleted or renamed.

Nevertheless, platforms like GitHub are important for open-source software, as they offer a suite of features designed to foster an open and collaborative environment. 
Some of these features, however, are in direct conflict with open science principles. 
For example, GitHub allows deleting a repository.
However, already published articles are difficult to update, such that links to research artifacts may no longer be resolved.
This undermines the recoverability of the artifact~\cite{minocher2020reproducibility}.
The conflict can be avoided by utilizing both services jointly, i.e., one archival service for artifact persistence and one Git-based version control service for collaboration.
Step-by-step guides to automatically archive GitHub project releases to Zenodo\footnote{\url{https://guides.github.com/activities/citable-code/}} and FigShare\footnote{\url{https://knowledge.figshare.com/articles/item/how-to-connect-figshare-with-your-github-account}} enable this synergy.

\subsubsection{Share}
\label{sec:guide:share}

Science entails dissemination as much as knowledge-building. 
Science without communication is as empty as science without findings, and neglecting the dissemination of research artifacts reduces their potential impact.
Sharing research artifacts is, hence, an integral part of managing them.
Social media platforms like Twitter lowered the barrier of dissemination significantly.
Announcing research articles by summarizing the motivation, approach, findings, and artifacts in a few sentences each is a valid first step to garner interest and invite interaction.

\subsection{Long-Term Vision}
\label{sec:guide:vision}

Our vision is to maintain the online GoogleDoc version of this guideline as our community's understanding of open science evolves, and the desires of our research community for open science standards grow.
This vision requires the collective knowledge and effort of our community.
Just as science is conducted, we hope to have this guideline discussed, reviewed, challenged, and updated.
For this reason, the GoogleDoc version has:
\begin{itemize}
    \item \textbf{Open access} for all to share, read, and comment (via the GoogleDocs comment feature)
    \item \textbf{Document notifications} enabled for the primary maintainers (for quick activity reaction)
    \item \textbf{Document history} and version numbers for transparent and traceable evolution
\end{itemize}
In addition, we archive significant increments via Zenodo for persistence.

\subsection{Projected Use}
\revmeta{
We believe that the guidelines presented in this section contribute to the availability, persistence, and usability of research artifacts in software engineering.
The guidelines support the authors of scientific work in preparing, disclosing, and sharing the artifacts connected to their research.
The checklist format in the guidelines~\mbox{\cite{montgomery2023artifact}} makes the content suitable also to junior researchers.

We further invite organizers of scientific events like conferences and workshops to disseminate the guidelines among the authors so that they can receive more systematic guidance when sharing their artifacts.
Instead of migrating artifact evaluation guidelines from one event website to the next, as they are typically used for only one instance of an event, our guidelines constitute a central and maintainable artifact that can be referred to from any website.
This shall ensure that any progress of the community regarding artifact availability is recorded centrally.
}

\section{Conclusion and Future Work}
\label{sec:conclusion}

The availability of research artifacts is a vital precondition for the reproducibility of scientific work~\cite{minocher2020reproducibility}, on which the reliability and robustness of scientific results hinges~\cite{anda2008variability,cacioppo2015social}. In this work, we contribute both to the availability of research artifacts of previous publications in the area of requirements quality research through a two-phase, crowd-sourced recovery initiative resulting in \datasetsimproved\ recovered data sets and \implementationsimproved\ recovered implementations and to the availability of research artifacts of future publications through the compilation of a concise, pragmatic artifact management guideline. Additionally, we derive insights into the reasons for unavailability based on our sample of 57 primary studies from the requirements quality literature, including empirical evidence for artifact availability improving over time and public hosting services positively influencing artifact persistence.

Improving the availability of research artifacts through adherence to open science principles is a continuous effort we aim to contribute to with this study. We hope the insights derived from our data analysis and the resources provided in our replication package, including the recovery request generation script and artifact management guideline, allow the reproduction of this study in other fields of research to extend the scope of this study and shape a more accessible landscape of research artifacts for future researchers.

\section*{Acknowledgements}
This work was supported by the KKS foundation through the S.E.R.T. Research Profile project at Blekinge Institute of Technology. 
We further thank the members of the NLP4RE community who supported the recovery initiative, especially Eduard Groen, Daniel Berry, Fabiano Dalpiaz, and Alessio Ferrari.

\bibliography{references}

\end{document}